\documentclass[aps,prb,superscriptaddress,reprint]{revtex4-1}
\usepackage{graphicx,siunitx,amssymb,natmove}
\usepackage[version=3]{mhchem}

\def\df{d_\text{f}}
\def\Ef{E_\text{f}}
\def\Jf{J_\text{f}}

\def\sigf{\sigma_\text{f}}

\def\Estor{E_\text{storage}}
\def\Eloss{E_\text{loss}}

\begin{document}

\title{Wide-Range Tunable Dynamic Property of Carbon Nanotube-Based Fibers}
\author{Jingna Zhao}
\affiliation{National Engineering Laboratory for Modern Silk, College of Textile and Clothing Engineering, Soochow
University, Ren'ai Road 199, Suzhou 215123, China}
\affiliation{Key Laboratory of Nano-Devices and Applications, Suzhou Institute of Nano-Tech and Nano-Bionics, Chinese
Academy of Sciences, Ruoshui Road 398, Suzhou 215123, China}
\author{Xiaohua Zhang}
\affiliation{Key Laboratory of Nano-Devices and Applications, Suzhou Institute of Nano-Tech and Nano-Bionics, Chinese
Academy of Sciences, Ruoshui Road 398, Suzhou 215123, China}
\author{Zhijuan Pan}
\affiliation{National Engineering Laboratory for Modern Silk, College of Textile and Clothing Engineering, Soochow
University, Ren'ai Road 199, Suzhou 215123, China}
\author{Qingwen Li}
\affiliation{Key Laboratory of Nano-Devices and Applications, Suzhou Institute of Nano-Tech and Nano-Bionics, Chinese
Academy of Sciences, Ruoshui Road 398, Suzhou 215123, China}

\begin{abstract}
Carbon nanotube (CNT) fiber is formed by assembling millions of individual tubes. The assembly feature provides the
fiber with rich interface structures and thus various ways of energy dissipation, as reflected by the non-zero loss
tangent ($>$0.028--0.045) at low vibration frequencies. A fiber containing entangled CNTs possesses higher loss tangents
than a fiber spun from aligned CNTs. Liquid densification and polymer infiltration, the two common ways to increase the
interfacial friction and thus the fiber's tensile strength and modulus, are found to efficiently reduce the damping
coefficient. This is because the sliding tendency between CNT bundles can also be well suppressed by the high packing
density and the formation of covalent polymer cross-links within the fiber. The CNT/bismaleimide composite fiber
exhibited the smallest loss tangent, nearly as the same as that of carbon fibers. At a higher level of the assembly
structure, namely a multi-ply CNT yarn, the inter-fiber friction and sliding tendency obviously influence the yarn's
damping performance, and the loss tangent can be tuned within a wide range, as similar to carbon fibers, nylon yarns, or
cotton yarns. The wide-range tunable dynamic properties allow new applications ranging from high quality factor
materials to dissipative systems.
\end{abstract}

\maketitle

%\section{Introduction}

Natural and synthetic fibers are used in many products and play a large role in everyday applications. Their mechanical
properties are important both from the point of view of fiber processing in a technological process and their use in the
form of final products. Besides the quasi-static properties including the fiber's tensile strength ($\sigf$), elastic
modulus ($\Ef$), and fracture toughness ($\Jf$), the dynamic response to a periodic external signal (displacement or
force) can be severe due to the energy dissipation \cite{menard.kp:1999}. Obviously, frictional effects play a major
role in the dynamic response. However, the types of friction differ greatly between a thread, yarn, and fiber due to
their different assembly structures.

Fiber is usually defined as a hair-like strand of material and is the smallest ``visible'' unit of a fabric. Natural
fibers (cotton, flax, jute, bamboo, kapok, ramie, wool, silk, and spider silk) and synthetic fibers (nylon, glass fiber,
cellulose fiber, and carbon fiber) usually have a width (diameter) ranging from several to tens of {\si \micro}m, and a
length at least 100 times longer than the width. Yarn is a continuous length of interlocked fibers used for
manufacturing textiles \cite{hearle.jws:20011}. At the uppermost structure level, threads are usually made by plying and
twisting yarns, for efficient and smooth stitching in sewn products.

When carbon nanotubes (CNTs) are assembled into a strand by a spinning process \cite{jiang.kl:2002, zhang.m:2004,
li.yl:2004, motta.m:2007, jia.jj:2011, zhang.xh:2012, lu.wb:2012}, a new man-made or synthetic fiber is formed. A CNT
fiber with a diameter of $\sim$10 \si{\micro m} usually contains more than $10^6$ individual CNTs. Such assembly feature
is different from those fibers well-known for many years which are usually a solid structure without internal
interfaces. CNT fiber can also be considered as a continuous length of interlocked ``filaments'' (CNT bundles)
\cite{zhao.jn:2010}, where the bundles are formed during the CNT growth rather than in the spinning process. Although
also being called as CNT yarn, CNT fiber is indeed not a yarn because the CNT bundles are not macroscopically
processable. On the contrary, the basic components of a yarn, the long and parallel or interlocked filaments, are
usually processable objects with a width larger than several \si{\micro m} \cite{hearle.jws:20011}.

As similar to the relationship between fiber and yarn, the friction between CNT bundles should also be the most
important way to dissipate energy in the dynamic responses of CNT fiber. In this paper a method for investigating
inter-bundle friction in CNT fiber is presented and the effect of interfacial enhancement on friction is shown. For the
pure CNT fibers, the higher level of densification, the lower loss tangent. The introduction of covalent polymer
networking can also reduce remarkably the damping phenomenon. For these two situations, although the increase in
frictional force results in higher ``localized'' energy dissipations, the ``overall'' sliding tendency between CNT
bundles can be remarkably suppressed due to the increased interfacial interactions and the covalent networks. However,
by tuning the type of CNT/polymer network from covalent to non-covalent and using different polymer chain lengths, the
loss tangent can be tailored within a wide range. Furthermore, the spinning CNT fibers into a multi-ply CNT yarn is
another important way to make them behavior like carbon fibers, nylon yarns, or cotton yarns. With considering that the
CNT fibers are flexible, electrically conductive, and mechanically strong, the tunable dynamic property can well extend
their applications ranging from high quality factor materials to dissipative systems.

%\section{Results and Discussion}

%\subsection{Analytical Development}

\begin{figure}[!t]
\centering
\includegraphics[width=.25\textwidth]{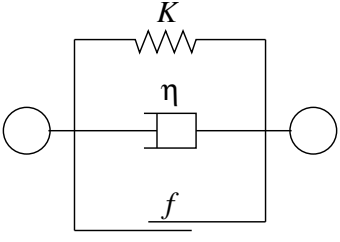}
\caption{\label{fig.kelvin-voigt} Schematic of a modified Kelvin-Voigt model where an elastic spring $K$, a viscous
damping coefficient $\eta$, and an inter-bundle friction $f$ are involved.}
\end{figure}

A modified Kelvin-Voigt model was adopted according to the studies on staple yarns \cite{murayama.t:1979}, where three
elements are connected in parallel to study the dynamic responses of CNT fibers (Figure \ref{fig.kelvin-voigt}). These
elements include a linear elastic spring $K$ which stores energy (i.e., the fiber's modulus $\Ef$), an energy
dissipation mechanism associated with the internal viscosity $\eta$ of the filaments or CNT bundles, and an energy
dissipation mechanism of the coulomb form associated with the sum of inter-bundle friction $f$. One should notice that
due to the hierarchical feature $\eta$ is also related to the inter-tube friction within a CNT bundle.

When a sinusoidal displacement $X=A \sin\omega t$ is applied, $A$ and $\omega$ being the amplitude and frequency, the
energy dissipations per cycle due to $\eta$ and $f$ are
\begin{gather}
\epsilon_\eta = \int \eta \frac{dx}{dt}dx = \eta A^2 \omega, \\
\label{eqn.4Af}
\epsilon_f = 4 \int_0^{\pi/2} f \frac{dx}{d(\omega t)}d(\omega t) = 4Af.
\end{gather}
By using an effective coefficient $\eta^* = \eta + 4f/A\omega$, the total energy dissipated can be represented by
$\epsilon_\eta+\epsilon_f = \eta^* A^2 \omega$.

The force output is
\begin{equation}
F=KX+\eta^*\frac{dx}{dt}=\frac{KA}{\cos\delta}\sin(\omega t+\delta),
\end{equation}
where
\begin{equation}
\label{eqn.tan}
\tan\delta=\eta^*\omega/K = \eta \omega / K + 4f/AK.
\end{equation}
This means that the effective damping loss tangent ($\tan\delta$) contains frequency- and friction-dependent components.
The traditional dynamic tests can just directly measure the effective loss tangent rather than the inter-bundle friction
$f$ or the bundle's internal viscosity $\eta$ for CNT fibers. However, by applying tests upon different $A$ and
$\omega$, it might be possible to study them separately, and in the present study the major interests have been focused
on the inter-bundle frictional properties.

%\subsection{Dynamic Tensile Tests}

The dynamic properties were measured with a Keysight T150 Universal Testing Machine under the Continuous Dynamic
Analysis (CDA) option, a direct and accurate way to measure the specimen's stiffness upon an additional vibrational load
input. As the displacement and force can be collected at each point in the experiment, it becomes possible to analyze
their amplitude vibrations and phase relationships, thus making it possible to determine the storage modulus ($\Estor$)
and loss modulus ($\Eloss$) \cite{keysight}. Notice that, due to the way to introduce vibrations in the T150 system, the
amplitude of vibration is controlled by the change in load force ($A_f$) rather than the displacement variation. The
default value $A_f= 4.5$ mN was used, which is less than 5\% of the fracture force for CNT fibers. Such value usually
resulted in a displacement amplitude of $A\approx3$--3.5 \si{\micro\m} for all the samples in the present study (see the
analysis in Experimental Section).

\begin{figure}[!t]
\centering
\includegraphics[width=.48\textwidth]{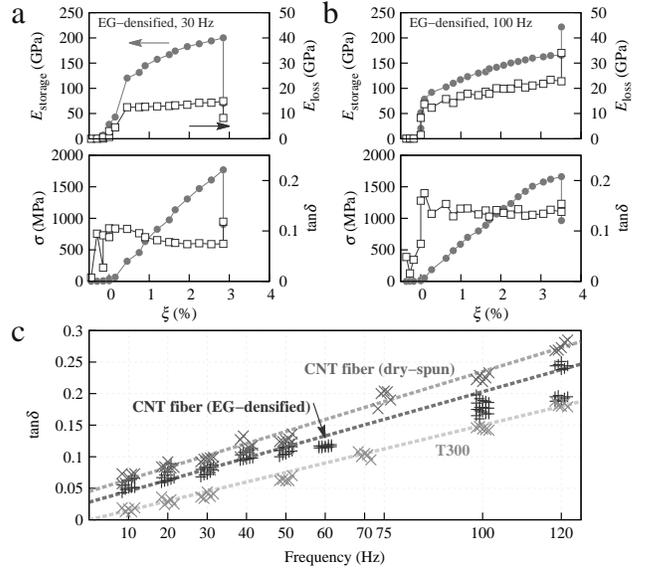}
\caption{\label{fig.tensile} Measurement of dynamic properties and their frequency dependence. (a,b) Instantaneous
$\Estor$, $\Eloss$, $\sigma$, and $\tan\delta$ for two tests on the EG-densified CNT fibers at 30 and 100 Hz,
respectively. (c) The frequency dependence of $\tan\delta$ for the dry-spun fiber, EG-densified fiber, and T300 carbon
fiber. Each data point corresponded to the average of an individual CDA tensile test.}
\end{figure}

Figure \ref{fig.tensile}a and b show two dynamic tests performed on ethylene glycol (EG)-densified CNT fibers at 30 and
100 Hz. Upon increasing the tensile strain ($\xi$), instantaneous engineering stress ($\sigma$), $\Estor$, and $\Eloss$
were obtained. The ratio between the loss and storage moduli, $\tan\delta = \Eloss/\Estor$, quickly increased up after
the beginning of stretching and was maintained until the fiber fracture. This provided an efficient way to measure the
dynamic property, i.e., by averaging the instantaneous values during the tensile test.

There existed a linear dependence of $\tan\delta$ on the frequency ($\omega$), see Figure \ref{fig.tensile}c, in
agreement with the analytical result (Equation \ref{eqn.tan}). However, by comparing to the dry-spun CNT fibers (see
Experimental Section) and Toray T300 carbon fibers, two remarkable differences can be observed. For the one hand, the
$\tan\delta$-to-$\omega$ slope for both types of CNT fibers was larger than that of the T300. On the other hand, by
extrapolating $\omega$ to zero, $\tan\delta$ at $\omega=0$ was non zero for the CNT fibers while it was nearly zero for
the T300. These two quantitative differences provided us a new way to study the frictional properties for CNT fibers.

The different $\tan\delta$-to-$\omega$ slopes mainly captured the difference in elastic modulus according to Equation
\ref{eqn.tan}. $\Ef$ was measured to be 204--238 GPa for the T300 carbon fiber (in agreement with their data sheet
\cite{t300}), while it was 45--59 GPa and 72--101 GPa for the dry-spun and EG-densified CNT fibers, respectively (see
also our previous results \cite{zhao.jn:2010, jia.jj:2011, li.s:2012}). As there should not be significant difference in
the intrinsic viscosity for the sp$^2$ structures in CNT and carbon fiber, the difference in $\Ef$ obviously became the
major source for the different linear $\tan\delta$-$\omega$ relationship. Furthermore, such difference also influences
greatly the quality factor, which might range from several hundred (for $\Ef\approx50$ GPa, like glass-reinforced
plastic \cite{wei.cy:2000}) to over 5000 (for $\Ef\approx200$--470 GPa, like carbon fiber
\cite{castellanos-gomez.a:2010} and sapphire fiber \cite{uchiyama.t:2000}). Thus to develop CNT fibers as high quality
factor materials, further improvement in modulus is of great necessity.

The non-zero extrapolated $\tan\delta$ at $\omega=0$ means that even without external vibrations, there always existed
energy losses at CNT interfaces. For the dry-spun and EG-densified CNT fibers $\tan\delta|_{\omega=0} = 0.045$ and
0.028, respectively. Such difference well reflected their different frictional feature in different assembly levels of
CNTs. Notice that, in a recent study, the passive damping of dry-spun CNT fibers exhibited energy loss ratio of 0.05
during a hysteresis test at 0.1 Hz \cite{hehr.a:2014}, in good agreement with our extrapolated results.

%\subsection{Assembly-Dependent Energy Dissipation}

To show the assembly dependence in more detail, various CNT-based assembly fibers were used. Besides the dry-spun and
EG-densified CNT fibers which were produced based on the array spinning method \cite{jiang.kl:2002, zhang.m:2004,
zhang.xb:2006, li.qw:2006}, some CNT fibers were obtained by directly spinning the aerogel-like CNT networks which were
grown with an injection chemical vapor deposition (iCVD) \cite{li.yl:2004, motta.m:2007, stano.kl:2008} and thus were
named iCVD fibers.

\begin{figure}[!t]
\centering
\includegraphics[width=.48\textwidth]{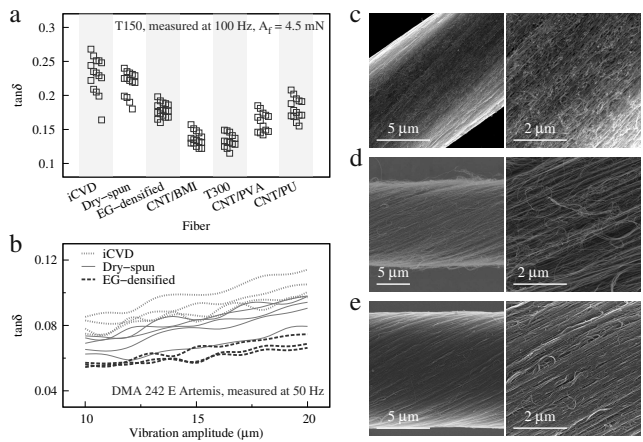}
\caption{\label{fig.comparison} Comparisons on $\tan\delta$ between various CNT fibers. (a) CNT assemblies have
remarkable influences on the damping property. (b) $\tan\delta$ increased slightly with vibration amplitude. (c-e) SEM
images showing the structure differences between the iCVD (c), dry-spun (d), and EG-densified (e) CNT fibers.}
\end{figure}

Figure \ref{fig.comparison}a shows the comparison on $\tan\delta$ for these different CNT fibers. The first attention
was paid to the three ``pure'' CNT fibers, namely the iCVD, dry-spun, and EG-densified fibers, where the CNT weight
percent was over at least 95\% \cite{li.s:2012, zhong.xh:2010}. The iCVD and dry-spun CNT fiber had the highest loss
tangents, averages of 0.229 and 0.218 at 100 Hz, respectively. When EG densification was introduced during the fiber
spinning \cite{li.s:2012}, the averaged $\tan\delta$ at 100 Hz decreased to 0.178. This difference can also be detected
with a DMA 242 E Artemis (Dynamic Mechanical Analyzer, see Experimental Section). By increasing the vibration amplitude
from 10 to 20 \si{\micro\m} at 50 Hz, the instantaneous $\tan\delta$ always exhibited the same sequence: iCVD $>$
dry-spun $>$ EG-densified (Figure \ref{fig.comparison}b).

It is necessary to point out that the vibration frequency in everyday life is usually not such high. Here we measured
$\tan\delta$ at 50 and 100 Hz just in order to show better comparisons as the data variation was small as compared with
its magnitude.

To analyze the assembly dependence, scanning electron microscopy (SEM) were performed for these CNT fibers, as shown in
Figure \ref{fig.comparison}c-e. For the iCVD CNT assemblies, the entangled CNT network (Figure \ref{fig.comparison}c)
contributed other ways of energy dissipation rather than the inter-bundle (inter-filament) friction. As reported in our
previous study \cite{liu.ql:2014}, the CNT network deformation was the main way to carry and transfer external loads and
the stress-strain curve clearly showed a wide plastic range, about from 2.5\% to $\sim$30\% in tensile strain. The
inelastic structural change at the CNT cross-links costed energies upon small vibrations and the slippage between CNTs
provided another energy-cost process at large strains. These structural changes at the cross-links were also named
detaching-attaching and zipping-unzipping processes \cite{li.y:2012}. On the contrary, the CNTs in the dry-spun fibers
were aligned and the number of CNT cross-links was much fewer than that in the iCVD network (Figure
\ref{fig.comparison}d). However, their packing density was not high and the CNT bundles can easily slide with each
other. Such assembly feature resulted in frictional energy consumption and thus the high loss tangent
\cite{koratkar.n:2002, koratkar.na:2003, suhr.j:2005}. After the EG densification, the improved packing density
increased the inter-bundle contact area and enhanced the interfacial frictional force. Such change would increase the
frictional energy dissipation according to Equation \ref{eqn.4Af}. However, the tendency of CNT slippage was remarkably
suppressed due to the high packing density, corresponding to the reduced number for summing up Equation \ref{eqn.4Af}.
Therefore, the overall energy dissipation was also suppressed.

Thus it is necessary to re-consider the content of Equation \ref{eqn.tan}. Accurately, $f$ describes the level of
overall frictional energy dissipation, reflecting not only the magnitude of frictional force but also the total number
of interfaces where sliding phenomenon might take place. Therefore, to develop CNT fibers as high quality factor or low
damping materials, the inter-bundle structure design becomes the key route.

%\subsection{Effect of Polymer Infiltration}

The tailoring on interface structure and interaction had effects not only on the damping property but also on the
quasi-static mechanical performances. For example, the liquid-induced high CNT packing density reduced the frictional
energy dissipation and also increased the tensile strength and modulus. The EG-densified CNT fibers demonstrated a
strength of 1.65--1.82 GPa and a modulus of 72--101 GPa (Figure \ref{fig.tensile}a and b), of $\sim$500 MPa and $\sim$40
GPa higher than those of the dry-spun fibers \cite{jia.jj:2011, li.s:2012}. This means that the strengthening methods
might also have significant effects on the tailoring of the dynamic property. For this purpose, polymer molecules such
as polyvinyl alcohol (PVA), polyurethane (PU), and bismaleimide (BMI) were infiltrated into CNT fibers as they all
improved the tensile properties significantly \cite{fang.c:2010, li.s:2012}.

The CNT/BMI composite fiber demonstrated the smallest loss tangent, which varied from 0.122 to 0.157 with an average of
0.136 at 100 Hz. Such value was very close to that for T300 (variation of 0.115--0.149 and average of 0.134), see Figure
\ref{fig.comparison}a. The infiltrated thermosetting BMI resins formed cross-lined polymer network within CNT fibers by
a curing process \cite{li.s:2012}, and such networking effect further suppressed the sliding tendency. By considering
the high tensile strength (up to 2.47 GPa) and high modulus (up to 110--140 GPa) \cite{li.s:2012, meng.fc:2014}, the
CNT/BMI composite fiber can be used as excellent high quality factor materials.

On the contrary, when thermoplastic polymers were introduced, although the inter-bundle interaction can be improved, the
viscosity of PVA or PU introduced new mechanisms to viscoelastically dissipate the energy \cite{yang.sy:2007}. Besides
this, the non-covalent interactions between the polymer molecule and CNT also resulted in network deformation upon
external strains. As a result, the measured $\tan\delta$ was higher than that of the CNT/BMI composite fiber
($\tan\delta \approx 0.161$ for the CNT/PVA and 0.18 for the CNT/PU, Figure \ref{fig.comparison}a). Notice that although
the interaction between CNT and BMI was non-covalent, the covalently bonded BMI network obviously was much more rigid
than the non-covalently entangled PVA and PU networks. The polymer chain length of BMI was also much shorter than that
of PVA and PU. These differences obviously resulted in different movability of CNT and thus the totally different
dynamic properties.

Therefore, for the highly packed CNT assemblies, the introduction of thermoplastic or thermosetting polymers can be an
efficient way to increase or decrease the loss tangent (while the tensile strength and modulus were both improved upon
the polymer infiltration \cite{li.s:2012}), and can be used for developing assembly fibers with a designed quality
factor.

%\subsection{Dynamic Property of Multi-Ply CNT Yarns}

\begin{figure}[!t]
\centering
\includegraphics[width=.35\textwidth]{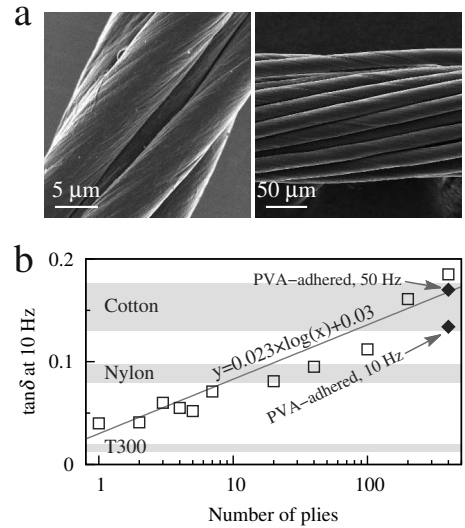}
\caption{\label{fig.multiply} Structure and loss tangent of multi-ply CNT yarns. (a) SEM images of a 7- and 40-ply CNT
yarn. (b) $\tan\delta$ at 10 Hz as a function of number of plies. For a better comparison, the results for T300 fibers,
nylon yarns, and cotton yarns are plotted as gray zones, and the results for a PVA-adhered 400-ply yarn at 10 and 50 Hz
are plotted as solid diamonds.}
\end{figure}

In textile industry, the spun agglomeration of fibers, named a yarn, is used for knitting, weaving, or sewing. The
simplest way to produce a CNT yarn is the multi-ply spinning. Figure \ref{fig.multiply}a shows the SEM images of a 7-ply
and 40-ply CNT yarn. Obviously, the multi-plying introduced the inter-fiber contacts as new sources of frictional energy
dissipation. By using the DMA 242 E Artemis with $A=10$ \si{\micro\m} and $f=10$ Hz, the dynamic properties of these
different yarns were characterized, as shown in Figure \ref{fig.multiply}b (the low frequency may often take place in
reality and the results can be compared to Figure \ref{fig.tensile}c). A logarithmic dependence of $\tan\delta =
0.023\times\log(n) + 0.03$ was observed with increasing the number of plies ($n$), see Figure \ref{fig.multiply}b.
$\tan\delta$ monotonically increased with $n$ because the inter-fiber contact areas increased to contribute more energy
dissipations. However, the dependence was not linear as the spinning treatment also hindered the sliding tendency
between CNT fibers. The simply multi-plying resulted in the highest loss tangent as shown in Figure \ref{fig.multiply}b,
e.g., $\tan\delta=0.185$ for the 400-ply yarn which had a linear density about 65 tex, that is, 65 g per kilometer. When
adhesion between the CNT fibers was also introduced in the multi-plying with polymer molecules, just like the enhanced
inter-bundle interaction with infiltrated polymers within a CNT fiber, the loss tangent also decreased slightly. For
example, the PVA-adhered 400-ply yarn exhibited a loss tangent of 0.134 at 10 Hz (Figure \ref{fig.multiply}b). Notice
that, for the multi-ply yarns, the similar frequency dependence of $\tan\delta$ was also observed. For example,
$\tan\delta$ increased up to 0.17 at 50 Hz for the PVA-adhered 400-ply yarn.

As a comparison, the loss tangents of cotton yarns, nylon 66 yarns, and T300 carbon fibers were also plotted in Figure
\ref{fig.multiply}b. The cotton yarns (linear density 55 tex) had the strongest damping performance, with
$\tan\delta|_{\omega=10 \text{ Hz}} = 0.13$--0.176 (our test results) or even higher than 0.2--0.3 from a recent dynamic
mechanical test (linear density 520 tex) \cite{qiu.yp:2001, qiu.y:2002}. The nylon yarns (linear density 80 tex)
exhibited a loss tangent of 0.08--0.098 at 10 Hz (in agreement with other literature reports \cite{demsar.a:2010}) as
their structure is more rigid than the cotton yarns. Different from these natural and synthetic yarns, the T300 fiber is
a structure even without an interface inside and thus had the weakest damping performance (notice that the carbon fiber
bundles did not show measurable increase in loss tangent as their high alignment and high modulus allowed the uniform
response to external strain or stress vibrations). Clearly, the multiplying on CNT fibers can efficiently tune the
dynamic property within a wide range as changing from carbon fiber to cotton yarn.

%\section{Conclusions}
CNT fiber is a new man-made synthetic high performance fiber and contains rich interfaces at the nanoscale. The assembly
feature introduces frictional energy dissipation and thus a high viscosity. The liquid densification increased the
inter-bundle contact and frictional force, but also remarkably suppressed the sliding tendency between CNT bundles. In
order to further suppress the sliding tendency, thermosetting polymer resins were infiltrated and cured to form covalent
networking within the CNT fiber. This resulted in a extremely low loss tangent, as small as that of Toray carbon fibers.
On the contrary, the introduction of thermoplastic polymers introduced new sources for energy dissipation. Finally, when
multiple CNT fibers were spun into an agglomeration, the multi-ply CNT yarn, the new inter-fiber contact also became
additional energy dissipative interfaces. With considering the high strength, modulus, and flexibility of CNT fibers, we
believe that this assembly can be developed as advanced low damping and high quality factor materials.

\section*{Experimental Section}

Most CNT fibers were produced by the array spinning method \cite{zhang.m:2004, zhao.jn:2010, jia.jj:2011}, where a CNT
sheet was pulled out from a vertically aligned CNT array and fabricated under drawing and twisting into a continuous
fiber. The twist angle was controlled to range from 15\si{\degree} to 20\si{\degree}, by controlling the ratio of
drawing and twisting speeds \cite{zhao.jn:2010}, and the fiber's diameter was 13--18 \si{\micro m}. The CNT arrays used
for fiber spinning were synthesized by a sustained chemical vapor deposition on \ce{SiO2}/Si wafers \cite{li.qw:2006},
with synthesis details reported elsewhere \cite{li.s:2012}. The grown CNTs were mainly double- to triple-walled and
$\leqslant$6 nm in diameter, see also our previous study \cite{meng.fc:2014}.

Various strengthening methods were reported for array-spun fibers, such as solvent densification and polymer
impregnation \cite{fang.c:2010, li.s:2012}. In the present study, the array spinning without using any solvent (also
called dry spinning), EG densification, PVA and PU infiltrations, and BMI infiltration and curing were used to produce
the dry-spun, EG-densified, CNT/PVA, CNT/PU, and CNT/BMI fibers. All the parameters for the infiltration treatment was
reported in our previous study \cite{li.s:2012}.

Some CNT fibers were produced by twisting and winding a thin CNT film drawn out of a surface where the injection CVD
growth was performed \cite{li.yl:2004, motta.m:2007, stano.kl:2008}, and thus were named iCVD fibers. The furnace had a
diameter of 40 mm. A mist of ethanol, ferrocene (2 wt\%), and thiophene (1 vol\%) was injected at a feeding rate of 0.15
ml/min. A gas mixture of Ar and \ce{H2} (volume ratio 1:1) was also flowed through the furnace at a rate of 4000 sccm.
The reaction was performed at 1300 \si{\celsius}. The iCVD fibers were mainly composed of double-walled CNTs.

The T150 Universal Testing Machine (Keysight Technologies Inc., Santa Rosa, USA) and the DMA 242 E Artemis
(NETZSCH-Ger\"{a}tebau GmbH, Selb, Germany) were used to characterize the dynamic properties. Frequencies ranging from
10 to 120 Hz were used in the T150 system, and 10--50 Hz in the DMA 242 E Artemis. The vibration amplitude was
controlled as load force variation and displacement variation in the two machines respectively. For the T150 system, the
default value of $A_f=4.5$ mN corresponded to a displacement amplitude of $A \approx$3--3.5 \si{\micro\m}. The
conversion requires the information of fiber diameter $\df$, elastic modulus $\Ef$, and gauge length $L$ for tensile
test, according to such relationship
\begin{equation}
A_f = \frac{\pi \df^2}{4} \Ef \frac{A}{L}.
\end{equation}
The temperature was controlled to be 20--25 \si{\celsius} in the DMA 242 E Artemis.

\section*{Acknowledgements}
The authors thank financial supports from the National Natural Science Foundation of China (21273269, 11302241,
11404371, 21473238) and Suzhou Industrial Science and Technology Program (ZXG201416).

%\end{thebibliography}
%\begin{small}
%\bibliographystyle{this}
%\bibliography{biblio,notes}

\begin{thebibliography}{10}

\bibitem{menard.kp:1999}
K.~P. Menard, editor, \emph{{Dynamic Mechanical Analysis: A Practical Introduction}},  (CRC Press, Boca Raton \textbf{1999}).

\bibitem{hearle.jws:20011}
J.~W.~S. Hearle, L.~Hollick, D.~K. Wilson, \emph{{Yarn Texturing Technology}}, (CRC Press, Boca Raton \textbf{2001}).

\bibitem{jiang.kl:2002}
K.~Jiang, Q.~Li, S.~Fan, \emph{Nature} \textbf{2002}, \emph{419}, 801.

\bibitem{zhang.m:2004}
M.~Zhang, K.~R. Atkinson, R.~H. Baughman, \emph{Science} \textbf{2004}, \emph{306}, 1358.

\bibitem{li.yl:2004}
Y.-L. Li, I.~A. Kinloch, A.~H. Windle, \emph{Science} \textbf{2004}, \emph{304}, 276.

\bibitem{motta.m:2007}
M.~Motta, A.~Moisala, I.~A. Kinloch, A.~H. Windle, \emph{Adv. Mater.} \textbf{2007}, \emph{19}, 3721.

\bibitem{jia.jj:2011}
J.~Jia, J.~Zhao, G.~Xu, J.~Di, Z.~Yong, Y.~Tao, C.~Fang, Z.~Zhang, X.~Zhang, L.~Zheng, Q.~Li, \emph{Carbon} \textbf{2011}, \emph{49}, 1333.

\bibitem{zhang.xh:2012}
X.~Zhang, Q.~Li, in Q.~Zhang, editor, \emph{{Carbon Nanotubes and Their Applications}}, chapter~14, pages 467--499, (Pan Stanford Publishing, Singapore \textbf{2012}).

\bibitem{lu.wb:2012}
W.~Lu, M.~Zu, J.-H. Byun, B.-S. Kim, T.-W. Chou, \emph{Adv. Mater.} \textbf{2012}, \emph{24}, 1805.

\bibitem{zhao.jn:2010}
J.~Zhao, X.~Zhang, J.~Di, G.~Xu, X.~Yang, X.~Liu, Z.~Yong, M.~Chen, Q.~Li, \emph{Small} \textbf{2010}, \emph{6}, 2612.

\bibitem{murayama.t:1979}
T.~Murayama, \emph{J. Appl. Polym. Sci.} \textbf{1979}, \emph{24}, 1413.

\bibitem{keysight}
a) Keysight T150 UTM Data Sheet: literature.cdn.key sight.com/litweb/pdf/5990-4206EN.pdf; b) Continuous Dynamic
Analysis Option - Data Sheet: literature.cdn.key sight.com/litweb/pdf/5990-4207EN.pdf; c) Continuous Dynamic
Analysis and Quasi-Static Measurement of Spider: literature.cdn.keysight.com/litweb/pdf/5990-4325EN.pdf.

\bibitem{t300}
TORAYCA carbon fibers T300 data sheet: www.toraycfa. com/pdfs/T300DataSheet.pdf.

\bibitem{li.s:2012}
S.~Li, X.~Zhang, J.~Zhao, F.~Meng, G.~Xu, Z.~Yong, J.~Jia, Z.~Zhang, Q.~Li, \emph{Compos. Sci. Technol.} \textbf{2012}, \emph{72}, 1402.

\bibitem{wei.cy:2000}
C.~Y. Wei, S.~N. Kukureka, \emph{J. Mater. Sci.} \textbf{2000}, \emph{35}, 3785.

\bibitem{castellanos-gomez.a:2010}
A.~Castellanos-Gomez, N.~Agra\"{i}t, G.~Rubio-Bollinger, \emph{Nanotechnology} \textbf{2010}, \emph{21}, 145702.

\bibitem{uchiyama.t:2000}
T.~Uchiyama, T.~Tomaru, D.~Tatsumi, S.~Miyoki, M.~Ohashi, K.~Kuroda, T.~Suzuki, A.~Yamamoto, T.~Shintomi, \emph{Phys. Lett. A} \textbf{2000}, \emph{273}, 310.

\bibitem{hehr.a:2014}
A.~Hehr, M.~Schulz, V.~Shanov, A.~Song, \emph{J. Intelligent Mater. Systems Structures} \textbf{2014}, \emph{25}, 713.

\bibitem{zhang.xb:2006}
X.~Zhang, K.~Jiang, C.~Feng, P.~Liu, L.~Zhang, J.~Kong, T.~Zhang, Q.~Li, S.~Fan, \emph{Adv. Mater.} \textbf{2006}, \emph{18}, 1505.

\bibitem{li.qw:2006}
Q.~Li, X.~Zhang, R.~F. DePaula, L.~Zheng, Y.~Zhao, L.~Stan, T.~G. Holesinger, P.~N. Arendt, D.~E. Peterson, Y.~T. Zhu, \emph{Adv. Mater.} \textbf{2006}, \emph{18}, 3160.

\bibitem{stano.kl:2008}
K.~L. Stano, K.~Koziol, M.~Pick, M.~S. Motta, A.~Moisala, J.~J. Vilatela, S.~Frasier, A.~H. Windle, \emph{Int. J. Mater. Form.} \textbf{2008}, \emph{1}, 59.

\bibitem{zhong.xh:2010}
X.-H. Zhong, Y.-L. Li, Y.-K. Liu, X.-H. Qiao, Y.~Feng, J.~Liang, J.~Jin, L.~Zhu, F.~Hou, J.-Y. Li, \emph{Adv. Mater.} \textbf{2010}, \emph{22}, 692.

\bibitem{liu.ql:2014}
Q.~Liu, M.~Li, Y.~Gu, Y.~Zhang, S.~Wang, Q.~Li, Z.~Zhang, \emph{Nanoscale} \textbf{2014}, \emph{6}, 4338.

\bibitem{li.y:2012}
Y.~Li, M.~Kr\"{o}ger, \emph{Soft Matter} \textbf{2012}, \emph{8}, 7822.

\bibitem{koratkar.n:2002}
N.~Koratkar, B.~Wei, P.~M. Ajayan, \emph{Adv. Mater.} \textbf{2002}, \emph{14}, 997.

\bibitem{koratkar.na:2003}
N.~A. Koratkar, B.~Wei, P.~M. Ajayan, \emph{Compos. Sci. Technol.} \textbf{2003}, \emph{63}, 1525.

\bibitem{suhr.j:2005}
J.~Suhr, N.~Koratkar, P.~Keblinski, P.~Ajayan, \emph{Nat. Mater.} \textbf{2005}, \emph{4}, 134.

\bibitem{fang.c:2010}
C.~Fang, J.~Zhao, J.~Jia, Z.~Zhang, X.~Zhang, Q.~Li, \emph{Appl. Phys. Lett.} \textbf{2010}, \emph{97}, 181906.

\bibitem{meng.fc:2014}
F.~Meng, X.~Zhang, R.~Li, J.~Zhao, X.~Xuan, X.~Wang, J.~Zou, Q.~Li, \emph{Adv. Mater.} \textbf{2014}, \emph{26}, 2480.

\bibitem{yang.sy:2007}
S.~Yang, J.~Taha-Tijerina, V.~Serrato-Diaz, K.~Hernandez, K.~Lozano, \emph{Compos. Part B} \textbf{2007}, \emph{38}, 228.

\bibitem{qiu.yp:2001}
Y.~Qiu, Y.~Wang, J.~Z. Mi, M.~A. Laton, X.~Shao, C.~Zhang, in \emph{{National Textile Center Annual Report}} (\textbf{2001}) page F98-S09.

\bibitem{qiu.y:2002}
Y.~Qiu, Y.~Wang, M.~Laton, J.~Z. Mi, \emph{Text. Res. J.} \textbf{2002}, \emph{72}, 585.

\bibitem{demsar.a:2010}
A.~Dem\v{s}ar, V.~Buko\v{s}ek, A.~Kljun, \emph{Fibres Text. East. Eur.} \textbf{2010}, \emph{18}, 29.

\end{thebibliography}
%\end{small}

\end{document}